\documentclass[aps,twocolumn,prl,showpacs,tightenlines]{revtex4}
\usepackage{amssymb}
\usepackage{epsfig}
\usepackage{amsbsy}
\usepackage{amsmath}
\usepackage{graphicx}
\usepackage{txfonts}

\begin{document}


\title{Two-Photon Scattering in One Dimension by Localized Two-Level System}
\author{T. Shi}
\affiliation{Institute of Theoretical Physics, Chinese Academy of
Sciences, Beijing 100190, China}
\author{C. P. Sun}
\affiliation{Institute of Theoretical Physics, Chinese Academy of
Sciences, Beijing 100190, China}

\begin{abstract}
We study two-photon scattering in a one-dimensional coupled
resonator arrays (CRA) by a two-level system (TLS), which is
localized as a quantum controller. The $S$-matrix is analytically
calculated for various two-photon scattering processes by TLS,
e.g., one photon is confined by TLS to form a bound state while
the other is in the scattering state. It is discovered from the
poles of the $S$-matrix that there exist two kinds of three-body
bound states for describing two bound photons localized around
TLS.
\end{abstract}

 \maketitle

\textit{Introduction.}---For the architecture of a new generation
all-optical quantum devices, we need to study various physical mechanisms of
the coherent photon transports in the low dimensional confined structure
with sub-wavelength scale. Recently, the single photon transmissions in a
waveguide and one dimensional (1D) coupled resonator arrays (CRA) with some
controllers, e.g., a two-level system (TLS), have been extensively studied~%
\cite{Fan1,Zhou,Xu,Shi,Fan3}. It is worthy to notice that, in such CRA
hybrid system, there usually exist bound states of single photon around the
localized TLS~\cite{Xu,Shi}, thus TLS can behave as a quantum controller to
realize a quantum switch~\cite{quantum switch} for photon transports.

We emphasize that most of investigations in this area focused on the cases
with single photon, and two-photon case was only considered for the linear
dispersion relation of waveguides~\cite{Fan1,Shi,Fan2}. In this Letter, we
develop a two-photon scattering approach for the general case with
arbitrarily given nonlinear dispersion relation. Actually, the effects of
TLS on the quantum natures of multi-photon statistics, such as photon
bunching and anti-bunching, have not been investigated systematically for
the nonlinear dispersion relations. However, a comprehensive understanding
for fundamental processes of multi-photon scattering is necessary to both
practical applications and theoretical explorations. In experiments, the
practical processes may concern two or more photons, which obviously affect
on the efficiency of single photon creation and transfer. In theoretical
studies, the hybrid system with many photons is related to the Lee model~%
\cite{Lee} in the sector with high excitations.

\begin{figure}[tbp]
\includegraphics[bb=57 399 570 768, width=8 cm, clip]{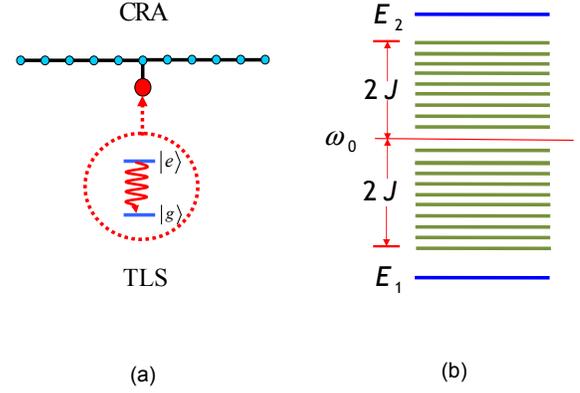}
\caption{(Color online) (a) The schematic for the CRA system: the red circle
denotes the two level system (TLS), and the blue dots denote the coupled
resonators. (b) The band structure of the CRA system for the single
excitation case: the continuum displays the single photon scattering states
while the blue lines above the band top and below the band bottom represent
the single photon bound states.}
\label{fig1}
\end{figure}

Our hybrid system consists of the CRA system (Fig.~\ref{fig1}a) with a
nonlinear dispersion relation~\cite{Zhou,Xu,Shi} and a TLS localized in one
of the resonators. It has been shown that in some circumstances there exist
two single-photon bound states, and energy band structure for a scattered
photon, which are displayed in Fig.~\ref{fig1}b. With this illustration, we
address our main results: (a) when scattered by TLS, which simultaneously
binds another photon to form a bound state, the incident photon is
elastically scattered under a condition we will discuss as follows; (b) when
two incident photons are scattered by TLS, the photon correlation is
induced, and the nonlinear dispersion results in richer quantum statistical
characters of out-going two photons, which can be controlled by TLS; (c)
there exist two kinds of three-body bound states for describing two bound
photons localized around TLS.

\textit{Model.---}The model Hamiltonian of our hybrid system reads%
\begin{equation}
H=\Omega \left\vert e\right\rangle \!\left\langle e\right\vert +\int
dk\varepsilon _{k}a_{k}^{\dagger }a_{k}+V\int \frac{dk}{\sqrt{2\pi }}%
(a_{k}^{\dagger }\sigma ^{-}+\text{\textrm{H.c.}})\,,  \label{HT}
\end{equation}%
where the operator $\sigma ^{-}=\left\vert g\right\rangle \left\langle
e\right\vert $ denotes the flip from the ground state (GS) $\left\vert
g\right\rangle \ $to the excited state $\left\vert e\right\rangle $ with the
energy level spacing $\Omega $. The nonlinear dispersion relation of photons
in CRA is $\varepsilon _{k}=\omega _{0}-2J\cos k$, where $\omega _{0}$ is
the eigen-frequency of each cavity, $J$ is the hopping constant\
characterizing the inter-cavity coupling in the tight-binding approximation,
$k$ is the momentum of photon, and the inter-cavity distance is taken as
unity. Here, $a_{k}\,$($a_{k}^{\dagger }$) is the annihilation$\,$(creation)
operator for the photon with momentum $k$, and $V$ is the coupling constant
of TLS and photon.

In the single excitation subspace spanned by the basis $\{\left\vert
0\right\rangle \left\vert e\right\rangle ,a_{k}^{\dagger }\left\vert
0\right\rangle \left\vert g\right\rangle \}$~\cite{Xu,Shi}, the system
possesses an energy band\ of width $4J$\ centered in $\omega _{0}$ and two
single-photon\ bound states $\left\vert E_{s}\right\rangle =\sqrt{Z_{s}}%
[\left\vert 0\right\rangle \left\vert e\right\rangle +V(2\pi )^{-1/2}\int
dk(E_{s}-\varepsilon )^{-1}a_{k}^{\dagger }\left\vert 0\right\rangle
\left\vert g\right\rangle ]$ $(s=1,2)$ with energies $E_{1}$\ below the band
bottom $\varepsilon _{\min }=\omega _{0}-2J$ and $E_{2}$ above the band top $%
\varepsilon _{\max }=\omega _{0}+2J$ (Fig.~\ref{fig1}b). Here, the
normalization constant is $Z_{s}^{-1}=1+V^{2}\int dk(E_{s}-\varepsilon
_{k})^{-2}/2\pi $ ($s=1,2$). $E_{s}$\ are two real\ solutions of $%
G^{-1}(E_{s})=0$~\cite{Xu,Shi}, where $G(\omega )$ is the Green's function
of photon, and $G^{-1}(\omega )=\omega -\Omega -\Sigma (\omega )$.\ Here, $%
\Sigma (\omega )=V^{2}\int dk(\omega -\varepsilon _{k})^{-1}/2\pi $ is the
self energy of photon.

\begin{figure}[tbp]
\includegraphics[bb=32 344 570 683, width=8 cm, clip]{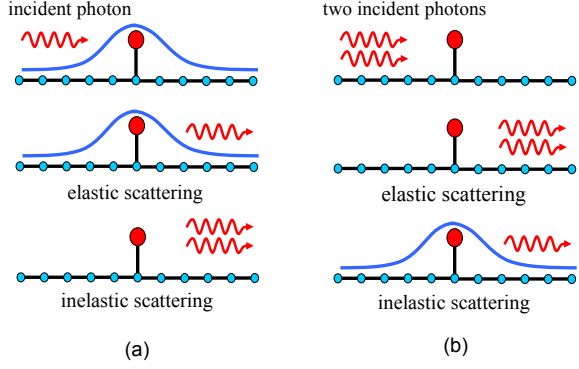}
\caption{(Color online) (a) The schematic for the photon scattering by TLS
in the single photon bound state: the first line indicates the single photon
incident, the second line means the scattering into the out-going state of
single photon, and the third line shows that the bound state is broken so
that there are two out-going photons; (b) the schematic for two-photon
scattering by the TLS in GS: the first line indicates the two photon
incident, the second line means the scattering into two out-going photons,
and the third line shows that, while a bound state form, another photon is
scattered into out-going state. }
\label{fig2}
\end{figure}

For two-photon scattering, there may exist three possible processes: (a) one
incident photon with momentum $k$ scattered by TLS in the bound state $%
\left\vert E_{1}\right\rangle $ or $\left\vert E_{2}\right\rangle $ (see
Fig.~\ref{fig1}a); (b) two incident photons scattered by TLS in GS; (c) two
photons both bound by TLS form the three-body bound state. We point out that
the behaviors of two photons in CRA are essentially different from those of
photons propagating in a waveguide, in which the processes (a) and (c) do
not exist due to its linear dispersion relation. The processes (a) and (b)
are schematically shown in Fig.~\ref{fig2}a and \ref{fig2}b, respectively.

\textit{Scattering eigen-states.---}The scattering eigen-states for the
two-photon processes are assumed to be
\begin{equation}
\left\vert \Psi ^{(l)}\right\rangle =\int dk\Phi _{e}^{(l)}(k)a_{k}^{\dagger
}\left\vert 0\right\rangle \left\vert e\right\rangle +\int dkdk^{\prime
}\Phi ^{(l)}(k,k^{\prime })a_{k}^{\dagger }a_{k^{\prime }}^{\dagger
}\left\vert 0\right\rangle \left\vert g\right\rangle
\end{equation}%
for $l=1,2,3$. The corresponding eigenvalues are $E^{(l)}=E_{l}+\varepsilon
_{k_{0}}$ for $l=1,2$ and $E^{(3)}=\varepsilon _{k_{1}}+\varepsilon _{k_{2}}$%
. Here, the eigen-states $\left\vert \Psi ^{(1,2)}\right\rangle $ describe
one photon of momentum $k_{0}$ scattered by TLS in the bound state $%
\left\vert E_{1,2}\right\rangle $, while $\left\vert \Psi
^{(3)}\right\rangle $ describes the two photons with momenta $k_{1}$ and $%
k_{2}$ scattered by TLS in GS. For convenient, we set $\varepsilon
_{0}=\varepsilon _{k_{0}}$, $\varepsilon _{i}=\varepsilon _{k_{i}}$, $%
\varepsilon =\varepsilon _{k}$ and $\varepsilon ^{\prime }=\varepsilon
_{k^{\prime }}$ below.

According to the three-particle scattering approach~\cite{A,AP,GT,S} for the
Lee model~\cite{Lee}, the solutions of scattering eigen-states for the
secular equation $H\left\vert \Psi ^{(s)}\right\rangle =E^{(s)}\left\vert
\Psi ^{(s)}\right\rangle $ ($s=1,2$) are obtained as%
\begin{equation}
\Phi ^{(s)}(k,k^{\prime })=\frac{V[\Phi _{e}^{(s)}(k)+\Phi
_{e}^{(s)}(k^{\prime })]}{2\sqrt{2\pi }(E^{(s)}-\varepsilon -\varepsilon
^{\prime }+i0^{+})},
\end{equation}%
where $\Phi _{e}^{(s)}(k)=\sqrt{Z_{s}}[\delta _{kk_{0}}+\psi (\varepsilon
,\varepsilon _{0})]$ contains the incident component $\delta _{kk_{0}}$ in
the $k$-space. The scattering part for single photon reads as%
\begin{eqnarray}
\psi (\varepsilon ,\varepsilon _{0}) &=&-\frac{V^{2}}{2\pi (\varepsilon
_{0}-\varepsilon )}[\frac{\varepsilon _{0}-E_{s}}{\varepsilon -E_{s}}%
G_{+}(\varepsilon _{0})  \notag \\
&&+\frac{2A_{s}(\varepsilon ,\varepsilon _{0})G_{+}(\varepsilon _{0})}{%
Z_{s}G_{+}(\varepsilon _{0})-A_{s}(\varepsilon _{0},\varepsilon _{0})}],
\end{eqnarray}%
where $A_{s}(\varepsilon ,\varepsilon _{0})=Z_{\bar{s}}(E_{\bar{s}%
}-E_{s})G_{+}(E^{(s)}-E_{\bar{s}})/(E^{(s)}-E_{\bar{s}}-\varepsilon
)+\int_{\varepsilon _{\min }}^{\varepsilon _{\max }}d\varepsilon ^{\prime
}J_{+}(E^{(s)}-\varepsilon ^{\prime },\varepsilon )(\varepsilon ^{\prime
}-E_{s})$Im$G_{+}(\varepsilon ^{\prime })$, and we define $%
F_{+}(x)=F(x+i0^{+})$ for arbitrarily given function$\ F(x)$. Here, $%
J(E,\varepsilon )=\pi ^{-1}G(E)/(\varepsilon -E)$ and $\bar{s}\neq s$ ($\bar{%
s}=1,2$).

For two-photon scattering by TLS in GS, the two-photon wave-function%
\begin{equation}
\Phi ^{(3)}(k,k^{\prime })=\delta _{kk_{1}}\delta _{k^{\prime }k_{2}}+\frac{%
V[\Phi _{e}^{(3)}(k)+\Phi _{e}^{(3)}(k^{\prime })]}{2\sqrt{2\pi }%
(E^{(3)}-\varepsilon -\varepsilon ^{\prime }+i0^{+})}
\end{equation}%
contains the incident two-photon component $\delta _{kk_{1}}\delta
_{k^{\prime }k_{2}}.$ and the two-photon wave-functions in the scattering
part%
\begin{equation}
\Phi _{e}^{(3)}(k)=\frac{V}{(2\pi )^{3/2}}[V^{2}\phi (\varepsilon
)G_{+}(\varepsilon _{1})G_{+}(\varepsilon _{2})+2\pi
G_{+}(E^{(3)}-\varepsilon )\sum_{i}\delta _{kk_{i}}],
\end{equation}%
where $\phi (\varepsilon )=2I(\varepsilon )/A-\sum_{i=1,2}\left( \varepsilon
-\varepsilon _{i}\right) ^{-1}$. Here, $I(\varepsilon
)=\sum_{s=1,2}Z_{s}G_{+}(E^{(3)}-E_{s})/(E^{(3)}-E_{s}-\varepsilon
)+\int_{\varepsilon _{\min }}^{\varepsilon _{\max }}d\varepsilon ^{\prime
}J_{+}(E^{(3)}-\varepsilon ^{\prime },\varepsilon )$Im$G_{+}(\varepsilon
^{\prime })$,$\ A=C-\sum_{s=1,2}Z_{s}G_{+}(E^{(3)}-E_{s})$, and $C=\pi
^{-1}\int_{\varepsilon _{\min }}^{\varepsilon _{\max }}d\varepsilon ^{\prime
}G_{+}(E^{(3)}-\varepsilon ^{\prime })$Im$G_{+}(\varepsilon ^{\prime })$.

\textit{Scattering by TLS in bound states.---}For\textit{\ }an incident
photon with momentum $k_{0}$ scattered by TLS in the bound state $\left\vert
E_{s}\right\rangle $, the out-going state $\left\vert \mathrm{out}%
\right\rangle $ contains the single photon component $a_{p_{0}}^{\dagger
}\left\vert E_{s}\right\rangle $ and the two-photon out-going part $%
a_{p_{1}}^{\dagger }a_{p_{2}}^{\dagger }\left\vert 0\right\rangle \left\vert
g\right\rangle $. The above obtained scattering eigen-states $\left\vert
\Psi ^{(1.2)}\right\rangle $ result in the $S$-matrix~\cite{Wick,Pauli,A,AP}%
. Its element%
\begin{equation}
S_{p_{0}k_{0}}^{(s)}=t_{k_{0}}^{(s)}\delta
_{k_{0}p_{0}}+r_{k_{0}}^{(s)}\delta _{-k_{0}p_{0}}  \label{S1}
\end{equation}%
represents an elastic process \textquotedblleft $\gamma +BS\rightarrow
\gamma +BS$\textit{\textquotedblright }: a photon $\gamma $ with momentum $%
k_{0}$ scattered by TLS in the bound state $\left\vert E_{s}\right\rangle $
into a photon $\gamma $ with momentum $p_{0}$ while the bound state is
unchanged. The transmission and reflection coefficients are $%
t_{k_{0}}^{(s)}=1+r_{k_{0}}^{(s)}$ and%
\begin{equation}
r_{k_{0}}^{(s)}=\frac{iV^{2}G_{+}(\varepsilon _{0})}{2J\left\vert \sin
k_{0}\right\vert }\frac{Z_{s}G_{+}(\varepsilon _{0})+A_{s}(\varepsilon
_{0},\varepsilon _{0})}{Z_{s}G_{+}(\varepsilon _{0})-A_{s}(\varepsilon
_{0},\varepsilon _{0})}.
\end{equation}

Another $S$-matrix element%
\begin{eqnarray}
S_{p_{1}p_{2}k_{0}}^{(s)} &=&i\frac{V^{3}}{\sqrt{\pi }}\sqrt{2Z_{s}}%
G_{+}(\varepsilon _{0})G_{+}(\varepsilon _{p_{1}})G_{+}(\varepsilon _{p_{2}})
\notag \\
&&\times \frac{\delta (E^{(s)}-\varepsilon _{p_{1}}-\varepsilon _{p_{2}})}{%
Z_{s}G_{+}(\varepsilon _{0})-A_{s}(\varepsilon _{0},\varepsilon _{0})}
\label{S2}
\end{eqnarray}%
means an inelastic process \textquotedblleft $\gamma +BS\rightarrow 2\gamma $%
\textit{\textquotedblright }: a photon $\gamma $ with momentum $k_{0}$
scattered by TLS in the bound state $\left\vert E_{s}\right\rangle $ into
two photons with momenta $p_{1}$ and $p_{2}$.

\begin{figure}[tbp]
\includegraphics[bb=0 162 593 700, width=8.5 cm, clip]{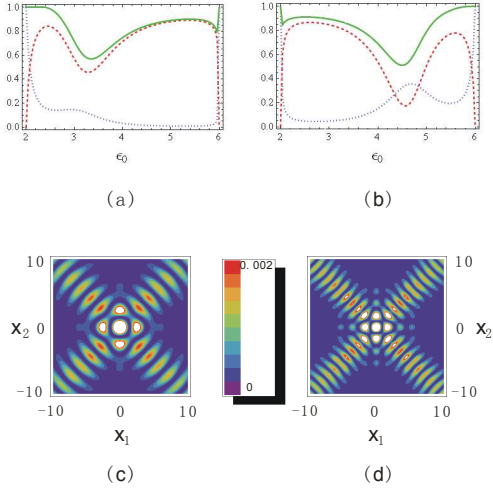}
\caption{(Color online) The transmission and reflection probabilities and
outgoing wave-function of two photons: the parameters of the system are $%
\protect\omega _{0}=4$ and $J=V=1$. In (a) and (c) $\Omega =2.5$, and in (b)
and (d) $\Omega =5$. The reflection probability, (blue dotted line)
transmission probability (red dashed line) and the sum (green solid line) of
them are drawn in (a) and (b) for the photon scattered by TLS in the lower
and higher bound states, respectively. The coordinate representation of two
photon out-going wave-functions are shown in (c) and (d) for photon with
energy $\protect\varepsilon _{0}=4.65$ scattered by TLS in the lower bound
state and photon with energy $\protect\varepsilon _{0}=2.55$ scattered by
TLS in the higher bound state, respectively.}
\label{fig3}
\end{figure}

Let us analyze the physical processes described by Eqs.~(\ref{S1}) and (\ref%
{S2}). When $\varepsilon _{0}<\varepsilon _{\mathrm{th}}^{(1)}=2\varepsilon
_{\min }-E_{1}$, $S_{p_{1}p_{2}k_{0}}^{(1)}$ vanishes, thus the incident
photon is elastically scattered by TLS in the bound state $\left\vert
E_{1}\right\rangle $. When $\varepsilon _{0}>\varepsilon _{\mathrm{th}%
}^{(1)} $, the bound state $\left\vert E_{1}\right\rangle $ is stimulated by
the incident photon, and the out-going two photons emerge simultaneously.
The incident photon with energy $\varepsilon _{0}>\varepsilon _{\mathrm{th}%
}^{(2)}=2\varepsilon _{\max }-E_{2}$ is elastically scattered by TLS in the
bound state $\left\vert E_{2}\right\rangle $ while the bound state $%
\left\vert E_{2}\right\rangle $ is not\ affected. When $\varepsilon
_{0}<\varepsilon _{\mathrm{th}}^{(2)}$ the out-going two photons emerge.

As shown in Fig.~\ref{fig3}a and Fig.~\ref{fig3}b, the reflective
probabilities $\left\vert r_{k_{0}}^{(s)}\right\vert ^{2}$ associated to\
the bound states are suppressed in comparison with that for the single
photon scattering in Ref.~\cite{Zhou}. Here, the probabilities $%
p_{k_{0}}=\left\vert r_{k_{0}}^{(s)}\right\vert ^{2}+\left\vert
t_{k_{0}}^{(s)}\right\vert ^{2}$ are also displayed in Fig.~\ref{fig3}a for $%
s=1$ and Fig.~\ref{fig3}b for $s=2$: when $\varepsilon _{0}<\varepsilon _{%
\mathrm{th}}^{(1)}$ ($\varepsilon _{0}>\varepsilon _{\mathrm{th}}^{(2)}$),
the probability $p_{k_{0}}=1$ means the elastic scattering by TLS in the
bound state $\left\vert E_{1}\right\rangle $ ($\left\vert E_{2}\right\rangle
$); when $\varepsilon _{0}>\varepsilon _{\mathrm{th}}^{(1)}$ ($\varepsilon
_{0}<\varepsilon _{\mathrm{th}}^{(2)}$), the probability $p_{k_{0}}<1$ means
the inelastic scattering by TLS in the bound state $\left\vert
E_{1}\right\rangle $ ($\left\vert E_{2}\right\rangle $), and in this case
the two photons emit. The probabilities $\left\vert \Phi
_{s}(x_{1},x_{2})\right\vert ^{2}$ for two-photon creation in the out-put,
which is defined by%
\begin{equation}
\Phi _{s}(x_{1},x_{2})=\int \frac{dp_{1}dp_{2}}{4\pi }%
S_{p_{1}p_{2}k_{0}}^{(s)}(e^{ip_{1}x_{1}+ip_{2}x_{2}}+e^{ip_{2}x_{1}+ip_{1}x_{2}}),
\end{equation}%
are plotted in Fig.~\ref{fig3}c and Fig.~\ref{fig3}d. The numerical
calculations show that the out-going two photons prefer to occupy around
TLS.
\begin{figure}[tbp]
\includegraphics[bb=0 162 595 731, width=8 cm, clip]{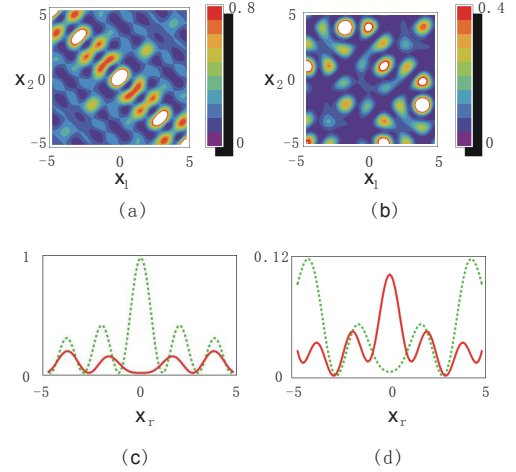}
\caption{(Color online) The second order correlation functions: the
parameters of system are $\Omega =\protect\omega _{0}=4,V=1$ and $J=1$, (a)
the second order correlation function for the two photons both in resonance
with TLS; (b) the second order correlation function for $\protect\varepsilon %
_{1}=3$ and $\protect\varepsilon _{2}=5$; (c) for $\protect\varepsilon _{1}=%
\protect\varepsilon _{2}=\Omega $, the dashed (Green) and solid (red) curves
denote the second order correlation function when $x_{c}$=0 and $x_{c}$=1,
respectively; (d) for $\protect\varepsilon _{1}=3$ and $\protect\varepsilon %
_{2}=5$, the dashed (Green) and solid (red) curves denote the second order
correlation function when $x_{c}$=0 and $x_{c}$=1.6, respectively.}
\label{fig4}
\end{figure}

\textit{Two-photon scattering by one TLS.---}For the two incident photons
both not bound by TLS, the out-going state $\left\vert \mathrm{out}%
\right\rangle $ generally contains the single-photon part $\left\vert
\mathrm{one}\right\rangle =\sum_{p,s=1,2}S_{pk_{1}k_{2}}^{(s)}a_{p}^{\dagger
}\left\vert E_{s}\right\rangle $ and the two-photon part $\left\vert \mathrm{%
two}\right\rangle
=\sum_{p_{1}p_{2}}S_{p_{1}p_{2}k_{1}k_{2}}a_{p_{1}}^{\dagger
}a_{p_{2}}^{\dagger }\left\vert 0\right\rangle \left\vert g\right\rangle $.
The Lippmann-Schwinger approach for $\left\vert \Psi ^{(3)}\right\rangle $
gives\ the $S$-matrix explicitly, which elements are $%
S_{pk_{1}k_{2}}^{(s)}=-iV^{3}W(2\pi Z_{s})^{-1/2}\delta (E_{s}+\varepsilon
_{p}-E^{(3)})$ ($s=1,2$) for the inelastic\ process \textquotedblleft $%
2\gamma \rightarrow \gamma +BS$\textit{\textquotedblright\ }and $%
S_{p_{1}p_{2}k_{1}k_{2}}=S_{\mathrm{in}}+R\delta (E^{(3)}-\varepsilon
_{p_{1}}-\varepsilon _{p_{2}})$ for the elastic process \textquotedblleft $%
2\gamma \rightarrow 2\gamma $\textit{\textquotedblright .\ }Here, $%
W=G_{+}(\varepsilon _{p})G_{+}(\varepsilon _{1})G_{+}(\varepsilon _{2})/A$,
and the first term $S_{\mathrm{in}%
}=S_{p_{1}k_{1}}S_{p_{2}k_{2}}+S_{p_{2}k_{1}}S_{p_{1}k_{2}}$ of $%
S_{p_{1}p_{2}k_{1}k_{2}}$ describes the factorization of the two-photon
scattering, where $S_{pk}=t_{k}\delta (p-k)+r_{k}\delta (p+k)$ contains the
transmission coefficient $t_{k}=1+r_{k}$ and reflection one%
\begin{equation}
r_{k}=-\frac{iV^{2}}{2J\left\vert \sin k\right\vert (\varepsilon _{k}-\Omega
)+iV^{2}}.
\end{equation}%
They describe the single photon scattering for TLS in GS~\cite{Zhou}. The
correlated photon scattering induced by TLS is described by the second term
of $S_{p_{1}p_{2}k_{1}k_{2}}$:%
\begin{equation}
R=-i\frac{V^{4}}{\pi A}\prod_{i=1,2}[G_{+}(\varepsilon
_{k_{i}})G_{+}(\varepsilon _{p_{i}})],
\end{equation}%
which\ the exhibits the background fluorescence of two photons~\cite%
{Fan1,Fan2,Shi}.

In the inelastic scattering of two photons, a part of incident photon energy
is absorbed by TLS, and thus the out-going state $\left\vert \mathrm{one}%
\right\rangle $ describes that one photon forms a bound state while another
emits. Another out-going state $\left\vert \mathrm{two}\right\rangle $ means
that the two photons are elastically scattered by TLS without any energy
loss. When $E_{1}+\varepsilon _{\max }<E^{(3)}<E_{2}+\varepsilon _{\min }$
the inelastic process is forbidden and only the elastic scattering takes
place. The second order correlation function~\cite{Shi} for two photons is $%
G(x_{1},x_{2})=\left\vert g(x_{1},x_{2})\right\vert ^{2}$, where%
\begin{equation}
g(x_{1},x_{2})=\int \frac{dp_{1}dp_{2}}{4\pi }%
S_{p_{1}p_{2}k_{1}k_{2}}(e^{ip_{1}x_{1}+ip_{2}x_{2}}+e^{ip_{2}x_{1}+ip_{1}x_{2}})
\end{equation}%
is the out-going wave-function in the coordinate representation.

Two correlation functions $G(x_{1},x_{2})$ are drawn in Fig.~\ref{fig4}a for
$\varepsilon _{1}=\varepsilon _{2}=\Omega $ and in Fig.~\ref{fig4}b for $%
\varepsilon _{1}=3$ and $\varepsilon _{2}=5$. In fact, due to the nonlinear
dispersion relation, the correlation function $G(x_{1},x_{2})$ not only
depends on the center-of-mass coordinate $x_{c}=(x_{2}+x_{1})/2$, but also
on the relative coordinate $x_{r}=x_{2}-x_{1}$ of two photons. For the
different $x_{c}$ the correlation functions $G$ are shown in Fig.~\ref{fig4}%
c for $\varepsilon _{1}=\varepsilon _{2}=\Omega $ and in Fig.~\ref{fig4}d
for $\varepsilon _{1}=3$ and $\varepsilon _{2}=5$. These figures indeed show
the different photon statistics for the different center-of-mass positions
of photons.

\begin{figure}[tbp]
\includegraphics[bb=4 441 590 710, width=8 cm, clip]{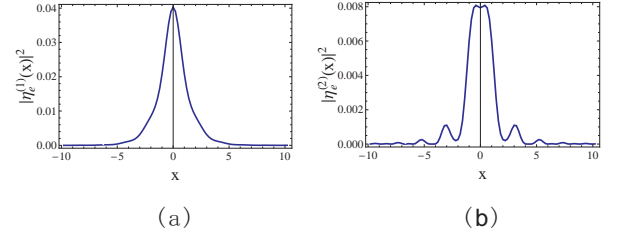}
\caption{(Color online) The wave-functions $|\protect\eta _{e}^{(s)}(x)|^{2}$
for $s=1,2$: the parameters of system are $\protect\omega _{0}=4$, $J=1$, $%
\Omega =3$, and $V=2$. As the photon goes to infinity, the probabilities
tend to zero.}
\label{fig5}
\end{figure}

\textit{Three-body bound states.---}It follows from the poles of $S$-matrix
that there exist two three-body bound states with energies $%
B_{s}=E_{s}+\beta _{s}$ ($s=1,2$), where $\beta _{s}$ are determined by $%
Z_{s}G_{+}(\beta _{s})=A_{s}(\beta _{s},\beta _{s})$. It can be proved that $%
\beta _{1}<\varepsilon _{\min }$ and $\beta _{2}>\varepsilon _{\max }$.
Using the approach in Ref.~\cite{BS}, we obtain the bound state%
\begin{equation}
\left\vert B_{s}\right\rangle =\mathcal{N}[\int dk\eta
_{e}^{(s)}(k)a_{k}^{\dagger }\left\vert 0\right\rangle \left\vert
e\right\rangle +\int dkdk^{\prime }\eta ^{(s)}(k,k^{\prime })a_{k}^{\dagger
}a_{k^{\prime }}^{\dagger }\left\vert 0\right\rangle \left\vert
g\right\rangle ]
\end{equation}%
by the secular equation $H\left\vert B_{s}\right\rangle =B_{s}\left\vert
B_{s}\right\rangle $. Here, $\eta _{e}^{(s)}(k)=V^{2}(\beta _{s}-\varepsilon
)^{-1}A_{s}(\varepsilon ,\beta _{s})/2\pi $,%
\begin{equation}
\eta ^{(s)}(k,k^{\prime })=\frac{V[\eta _{e}^{(s)}(k)+\eta
_{e}^{(s)}(k^{\prime })]}{2\sqrt{2\pi }(B_{s}-\varepsilon -\varepsilon
^{\prime })},
\end{equation}%
and $\mathcal{N}$ is the normalization constant. The wave-functions $\eta
_{e}^{(s)}(x)=\int dk\eta _{e}^{(s)}(k)\exp (ikx)/\sqrt{2\pi }$ and
\begin{equation}
\eta ^{(s)}(x_{1},x_{2})=\int \frac{dk_{1}dk_{2}}{2\pi }\eta
^{(s)}(k_{1},k_{2})(e^{ip_{1}x_{1}+ip_{2}x_{2}}+e^{ip_{2}x_{1}+ip_{1}x_{2}})
\end{equation}%
in the coordinate representation both tend to zero, when $x$, $x_{1}$ and $%
x_{2}$ tend to infinity, respectively. For the parameters $\omega _{0}=4$, $%
J=1$, $\Omega =3$, and $V=2$, the binding energies are estimated as$\
B_{1}\simeq 2.82$ and $B_{2}\simeq 12.54$, respectively. Here, $\left\vert
\eta _{e}^{(s)}(x)\right\vert ^{2}$ are shown in Fig.~\ref{fig5} for $s=1,2$%
, which both go to zero as $x\rightarrow \infty $.

\textit{Conclusion.---}We have investigated the two-photon scattering in
the\ hybrid CRA system. The two-photon $S$-matrix are calculated for various
scattering processes by TLS in the CRA system with nonlinear dispersion
relation. We show that the three-body bound states exist for describing the
two bound photons localized around TLS. The binding energies are determined
explicitly by the poles of $S$-matrix. For the two-photon scattering by TLS
in GS, the quantum statistics of scattered photon are analyzed in detail.
The similar approach can also be applied to dealing with scattering problems
of the TLS-photon hybrid system with\ more complicated architectures and
dispersion relations.

One (T. Shi) of the authors would like to thank Q. Ai for many helpful
discussion. The work is supported by National Natural Science Foundation of
China and the National Fundamental Research Programs of China under Grant
No. 10874091 and No. 2006CB921205.

\end{document}